\documentclass[5p]{elsarticle}

\usepackage{epsfig}
\usepackage{amsmath}
\usepackage{graphicx,psfrag}
\usepackage{dcolumn}
\usepackage{bm}
\usepackage{slashed}
\usepackage[dvipsnames]{xcolor}
\usepackage{amsfonts}
\usepackage{braket}

\newcommand{\beq}{\begin{equation}}
\newcommand{\eeq}{\end{equation}}
\newcommand{\bea}{\begin{eqnarray}}
\newcommand{\eea}{\end{eqnarray}}

\newcommand{\nn}{\nonumber\\}

\newcommand\fig[1]     {Fig.\,{\ref{#1}}}

\newcommand\app[1]   {~\ref{#1}}

\def\Ga{\Gamma}

\def\s0#1#2{\mbox{\small{$ \frac{#1}{#2} $}}}
\def\0#1#2{\frac{#1}{#2}}
\def\Eq#1{Eq.~(\ref{#1})}

\begin{document}

\title{From negative to positive cosmological constant through decreasing temperature of the Universe: 
connection with string theory and spacetime foliation results}

\author[1]{E.~N.~Nyergesy}

\author[2,1]{I.~G.~M\'ari\'an}

\author[3,4]{A.~Trombettoni}

\author[5,2]{I.~N\'andori}

\affiliation[1]{University of Debrecen, Institute of Physics, P.O.Box 105, H-4010 Debrecen, Hungary}
\affiliation[2]{HUN-REN Atomki, P.O.Box 51, H-4001 Debrecen, Hungary} 
\affiliation[3]{Department of Physics, University of Trieste, Strada Costiera 11, I-34151 Trieste, Italy}
\affiliation[4]{CNR-IOM DEMOCRITOS Simulation Center, Via Bonomea 265, I-34136 Trieste, Italy}
\affiliation[5]{University of Miskolc, Institute of Physics and Electrical Engineering, H-3515, Miskolc, Hungary}

\date{\today}

\begin{abstract}
String theories naturally predict a negative, while observations on the exponential expansion of the 
present Universe require a positive value for the cosmological constant.
Solution to resolve this discrepancy is known in the framework of string theory 
however, it might describe unstable worlds. Other options include modified 
$\Lambda$CDM models with sign switching cosmological constant (known as $\Lambda_s$ cosmology), 
but the sign flip is introduced into the models {\bf {\it ad hoc}}. Additional studies consider Asymptotically Safe (AS) 
quantum gravity by using Renormalization Group (RG), however their disadvantage is the omission of 
temperature which is otherwise crucial in the early Universe.
Here we present a proposal for resolving this conflict by using a modified thermal RG method where the 
temperature parameter $T$ is given by the inverse radius of the compactified time-like dimension, 
similarly to spacetime foliation. In our scenario not the dimensionful $T$, but the dimensionless  
temperature $\tau = T/k$ is kept constant when the RG scale $k$ is sent to zero and string 
theory is assumed to take place at very high while AS quantum gravity at intermediate 
and low temperatures. We show that the modified thermal RG study of AS quantum gravity 
models at very high temperatures results in a negative cosmological constant while turns it 
into a positive parameter for low temperatures.
\end{abstract}

\maketitle

\section{Introduction} 
In this work we present a proposal for resolving the conflict between string theories
and late time cosmology regarding the sign of the cosmological constant.
Indeed, the negative value of the cosmological constant was naturally predicted by string theories,
which contradicted the need for 
a positive value based on current cosmological observations. This 
motivated attempts to resolve 
the discrepancy in the framework of string theory \cite{kachru_2003} by wrapping antibranes,
but to get the required small and positive cosmological constant one has to wrap many fluxes
and it turned out that these solutions might describe unstable worlds \cite{hertog_2003}.

Motivated by observational data the $\Lambda_s$CDM model has been constructed 
\cite{lambdaS_cosmo_1, lambdaS_cosmo_2, lambdaS_cosmo_3} in which the cosmological 
constant $\Lambda$ of the $\Lambda$CDM model is replaced by a sign switching one ($\Lambda_s)$, 
i.e. $\Lambda \to \Lambda_s = \Lambda_{s0} \, \text{sgn}[z_\dagger - z]$, where $\Lambda_{s0} >0$ 
and $z_\dagger$ denotes the redshift at which the cosmological constant switches sign. They were 
able to predict such a $z_\dagger$ value in agreement with CMB+BAO data. However, the sign flip 
of $\Lambda$ is introduced artificially, which motivates the search for a phase transition that arises 
naturally within a model.

Additional attempts achieving the anti-de Sitter (AdS) -- de Sitter (dS) transition include considering 
two interacting dark energy fluids \cite{Sign_switch_1}, taking running Barrow entropy into account 
\cite{Sign_switch_2}, or examining quintessence fields with a negative cosmological constant 
\cite{Sign_switch_3}. Another example, for the AdS--dS transition is given in \cite{AdS_dS_1} where 
the generalization of the Hawking-Page phase transition \cite{Hawking_1983} is used.

The hypothetical AdS--dS transition -- which is said to alleviate the $H_0$ tension \cite{Hubble_1, Hubble_2, Hubble_3} 
and the $S_8$ discrepancy \cite{S8_1, S8_2, S8_3} -- is strongly influenced by the AdS Swampland conjecture 
\cite{Swampland_1,Swampland_2,Swampland_3} because the logarithmic dependence ($-\log |\Lambda|$) 
makes it impossible to cross the $\Lambda =0$ barrier at zero temperature \cite{AdS_dS_2, AdS_dS_3}. 
However, it was argued and shown in \cite{AdS_dS_2} that at finite temperature the number of light particles 
can change making the sign switch possible.

Further extensions for the $\Lambda$CDM model were constructed, 
e.g. the so called Dynamical Dark Energy (DDE) models \cite{DDE_1,DDE_2,DDE_3,DDE_4,DDE_5,DDE_6}
which can improve the joint consistency of supernova and BAO data. The sign switching $\Lambda$ models 
possess a constant $\omega = -1$ equation of state parameter, thus in a DDE scenario one would need 
an extended model to account for the change in $\omega$. However, in \cite{DDE_6} the possibility
whether the same data-driven preference for a decreasing dark energy density could be realized more 
naturally in sign-changing scenarios was mentioned.

Solution for the sign-problem of the cosmological constant has also been suggested in the framework
of quantum field theory (QFT) using a Renormalization Group (RG) in particular, the Functional
Renormalization Group (FRG) method \cite{frg_string}.
The concept was to assume string theory at the ultraviolet (UV) and Asymptotically Safe (AS) quantum 
gravity \& matter at the infrared (IR) scaling regimes connected by the RG scale $k$ which serves as 
a bridge between the negative UV and the positive IR values of the cosmological constant.
However, the RG scale $k$ is introduced artificially in the Wilsonian approach \cite{wilson}
and the quantized theory must be obtained in the physical limit $k \to 0$, so its use is not
fully justified to connect early and late time cosmologies. In addition,
temperature, which is 
a relevant parameter in cosmology, is missing in this approach.

Thus, here we suggest to use the temperature instead of the RG scale $k$ to connect string 
theory at very high temperature and AS quantum gravity at intermediate and low temperatures
of the expanding
Universe. To achieve this one has to extend the zero-temperature FRG method \cite{eea_rg}
to finite temperature. This can be done by using the standard finite temperature extension of QFT 
where the time-like dimension is compactified \cite{negele,sachdev,kapusta} and its inverse 
radius plays the role of the dimensionful temperature parameter $T$ (in natural units). 

In the usual finite-temperature perturbative RG approach the temperature parameter $T$ and
the running (perturbative) RG scale $\mu$ are linked to each other $\mu = 2\pi T$. The RG 
scales of the perturbative and the non-perturbative approaches are connected $\mu \sim k$. 
However, the $k$-dependence of the effective action of the non-perturbative FRG 
approach is introduced artificially to implement the Wilsonian integration of fluctuating modes and 
the quantised theory is obtained in the physical limit $k \to 0$ which has to be taken unavoidably.
Thus, in the usual finite temperature FRG approach, the dimensionful temperature 
parameter $T$ is linked to a fixed (not running) momentum scale $\Lambda$ which results in 
$T = \tau \Lambda$, see \cite{rg_dimful_T_1}. Most of the further literature on thermal FRG is 
based on this assumption. In this fixed $T$ approach, the dimensionful temperature is well defined 
and the limit $k \to 0$ can be done safely, but there is a price to pay: the RG flow equations have 
{\em no} non-trivial fixed point solutions. To overcome these difficulties, 
in Refs.~\cite{thermal_rg_cosmo,thermal_rg_as_gravity,thermal_rg_qpt_cpt} a modification of the 
usual FRG approach at finite temperature was proposed by relating the temperature parameter to 
the running RG scale as $T = \tau k$ (in natural units). In our approach
%
\begin{equation}
  \tau = T/k
  \label{tau_def}
\end{equation}
%
is kept constant over the RG flow while taking the simultaneous $T, k \to 0$ limit. Thus the 
dimensionful parameter $T$ changes by the RG scale.

Our working hypothesis is then to consider the dimensionful temperature parameter $T$
as a running cutoff for thermal fluctuation and to fix the parameter $\tau$ over the RG flow 
according \eqref{tau_def}: $T = \tau k$, where $T$ no longer plays the role of the temperature. 
We will use the notation $T \equiv k_T$ where $k_T$ is understood as the running cutoff for 
thermal and $k$ is for quantum fluctuations. In our approach the ``true'' temperature (the one 
entering the thermodynamic quantities such as the free energy that one aims at calculating) 
is proportional to the dimensionless quantity $\tau$ which is kept constant over the RG flow; 
it can be chosen arbitrarily, it is {\it not} restricted to a specific value and therefore can have 
any finite value. The relation between $\tau$ and the ``true'' temperature $T_{true}$ is
$\tau=k_B T_{true}/J$, where $J$ is an energy scale of the system depending on the microscopic 
details of the system. Since our results depend only on $\tau$ and not separately on $T_{true}$ 
and $J$, we keep the discussion in terms of $\tau$.  In other words, one can say $\tau$ measures 
how large the thermal fluctuations are compared to the quantum ones. Thus, our thermal RG 
scheme differs from the usual non-perturbative and also from the perturbative ones. The rest of 
the paper is devoted to understand the consequences of our working hypothesis on the thermal 
evolution of the cosmological constant.

\begin{figure}[t!] 
\begin{center} 
\epsfig{file=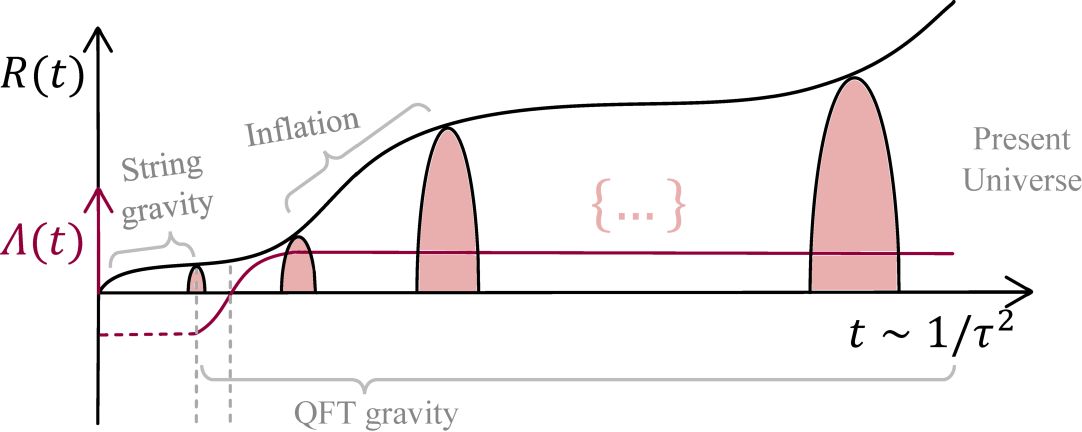,width=8.2cm}
\caption{\label{fig1} 
Schematic figure of the time evolution (temperature, i.e., $\tau$-dependence) of the scale factor 
$R(t)$ and the cosmological constant $\Lambda(t)$. 
} 
\end{center}
\end{figure}

In summary, in the fixed $T$ approach the dimensionful temperature is well 
defined, but RG flow equations have no fixed points. In our fixed $\tau$ approach the dimensionful 
temperature is not defined, but the fixed points are. This small, but crucial modification of the original 
thermal RG method opened the avenue to consider the interplay of classical (CPT) and
quantum (QPT) phase transitions, as discussed in \cite{thermal_rg_qpt_cpt} which gave us the 
possibility to compare it to simulation results and to confirm the viability of the fixed $\tau$ scheme.
Moreover, in the recent work \cite{thermal_rg_as_gravity} we studied the thermal RG of the 
simplest AS quantum gravity model, the Einstein-Hilbert (EH) truncation, using the fixed $\tau$
approach. 
The quantum effective action at a given dimensionless temperature
$\tau$ was given by moving along the thermal RG trajectory and this procedure was repeated 
for various values of $\tau$, which resulted in the $\tau$-evolution of the Reuter \cite{reuter} 
(i.e., non-Gaussian  UV) fixed point. We showed that in the high temperature limit ($\tau \to \infty$) 
the $g$-coordinate of the Reuter fixed point vanishes and the cosmological constant takes on a 
negative value in the limit $k\to 0$. It is not in disagreement with observations, since during the 
thermal evolution of the Universe a thermal phase transition occurs and the cosmological constant 
runs to the expected positive value at low temperatures. This mechanism to solve the sign-problem 
of the cosmological constant is represented on \fig{fig1}.
%
%

Motivated by the results of \cite{thermal_rg_as_gravity}, here we study the generality of these
findings. We find that the thermal RG study of basically any AS quantum gravity model  
(for recent reviews see \cite{AS_grav_1,AS_grav_2}) has the cosmological constant with a negative 
(positive, respectively) sign for large (small) temperature. Therefore, the scheme represented in 
\fig{fig1} can be used in every case which can be
seen as a natural solution for the sign problem of the cosmological constant. In particular, 
we study three extensions/variants of the simplest AS quantum gravity: 
the conformally reduced and the ghost-improved versions and its extension by scalar matter 
fields. In addition to that we discuss the connection to spacetime foliation.

\section{CREH gravity at $T=0$} 
As a first step, we summarise the main ideas behind the Conformally Reduced Einstein-Hilbert (CREH) 
truncated gravity at zero temperature 
\cite{CREH_Reuter_Weyer_2009}. In this model the approximation of 
Quantum Einstein Gravity (QEG) RG flow is done in two steps. Firstly,
one takes the usual EH truncation, then 
one performs the conformal reduction, where only the conformal factor of the metric is quantized.
The parametrization 
of the conformal factor is done in terms of the $\phi(x)$ scalar function
(with kinetic term $\sim (\partial_\mu \phi)^2$). 
In $d=4$ dimensions the metric is given as $g_{\mu \nu} = \phi ^2 \widehat{g}_{\mu \nu}$, where 
$\widehat{g}_{\mu \nu}$ is the non-dynamical reference metric. 
The EH action reads
\begin{equation}
S_{\rm EH}[g_{\mu \nu}] = \frac{1}{16 \pi G} \int d^4x \, \sqrt{g} \, (2\Lambda- \mathcal{R}(g)),
\end{equation}
where $G$ is the Newton constant, $\Lambda$ is the cosmological constant, $\mathcal{R}$ is the Ricci scalar 
and $g = \text{det}(g_{\mu \nu})$. Performing Weyl rescalings leads to a $\phi^4$-like theory
with $S_{\rm EH} = \int d^4x \, \sqrt{\widehat{g}}  {\cal L}_{\rm EH}$, with 
\begin{align}
\label{creh_action}
    {\cal L}_{\rm EH} = - \frac{3 }{4 \pi G} \, \bigg( \frac{1}{2} \widehat{g}^{\mu \nu} \partial_\mu \phi \partial_\nu \phi \, 
    + \frac{1}{12} \widehat{\mathcal{R}} \phi^2 - \frac{1}{6} \Lambda \phi^4  \bigg).
\end{align}
Since the kinetic term is negative in \Eq{creh_action}, a rapidly varying $\phi(x)$ could cause $S_{\rm EH}$ 
to become arbitrary negative, this being called the conformal factor instability. One can
introduce an inverted action 
($S_{\rm inv} \equiv -S_{\rm EH}$) to shift the negative sign to the potential term which is
used in the path integral. By using the RG formalism to the background field approach,
see \app{app1}, leads then to the effective 
action $\Ga_k[\bar{f}; \chi_B] =  \int d^4x \, \sqrt{\widehat{g}} {\cal L}^{\rm CREH}_k$ with
\begin{align}
{\cal L}^{\rm CREH}_k = - \frac{3}{4 \pi G_k} \left(-\frac{1}{2} \phi \widehat \Box  \phi  + \frac{1}{12} \widehat{\mathcal{R}} \phi^2 - \frac{1}{6} \Lambda_k \phi^4
\right),
\end{align}
where
$\widehat \Box = \widehat{g}^{-1/2} \partial_\mu \widehat{g}^{1/2} \widehat{g}^{\mu \nu} \partial_\nu$ 
denotes the Laplace-Bertrami operator belonging to the reference metric. In the RG flow equations 
dimensionless couplings are used, defined as
\begin{equation}
    g_k = k^2G_k, \hspace{1 cm} \lambda_k = k^{-2} \Lambda_k\,.
\end{equation}
leading to the following beta-functions 
\begin{align}
    \beta_g =& \, [\,2+\eta_N]\,g_k\,, \\
    \beta_\lambda =& \,-(2-\eta_N)\,\lambda_k 
    + \frac{g_k}{2\pi}\,[\,\Phi_2^1(-2\lambda_k)- \frac{1}{2}\,\eta_N\,\tilde{\Phi}_2^1(-2\lambda_k) \,], \nonumber
\end{align}
with threshold functions $\Phi_2^{1}(w)$, $\tilde{\Phi}_2^{1}(w)$ and anomalous dimension $\eta_N$, see \app{app1}.

\section{QEG with matter at $T=0$} 
We summarise briefly the RG study of QEG 
coupled to $N$-component scalar field at zero temperature \cite{Matter_Granda_1998}. The model
is 
interesting for the influence of the 
number of scalars and the scalar gravitational coupling on the flow of the Newtonian
and the cosmological constant. One has
${\cal S}_{\rm N} =   \int d^4x \sqrt{-g}  {\cal L}_{\rm N}$, with the Lagrangian density
\begin{align}
 {\cal L}_{\rm N} = \Bigg[ \frac{(-\mathcal{R} + 2 \Lambda) }{16\pi G} 
   + \frac{1}{2} \partial_\mu \phi^i \partial^\mu \phi_i  
    + \frac{1}{2} \xi \mathcal{R} \phi^i \phi_i \Bigg],
\end{align}
where $i= 1, \, ...,\, N$ and $\xi$ is the scalar gravitational coupling.
The background field method is used
during the derivation of the beta-functions, see \app{app1}, resulting in
\begin{align}
    \beta_g =& \, [\,2+ \eta_N(k)\,]\,g_k \,, \\
    \beta_\lambda =& -[\,2-\eta_N(k)\,]\lambda_k + \frac{1}{2 \pi} g_k \, [\,10 \, \Phi_2^1(-2\lambda_k)\, + \nonumber\\ 
    &+ \, (N-8)\, \Phi_2^1(0) -5\, \eta_N(k)\, \tilde{\Phi}_2^1(-2 \lambda_k)\, ]. \nonumber
\end{align}
%

\section{Ghost-improved EH gravity at $T=0$}
Lastly, let us study the EH truncated gravity at zero temperature with quantum effects captured 
by the wave-function renormalization $Z_k^c$, which multiplies the ghost-kinetic term.
One obtains the 
EH truncated gravity \cite{Reuter_2002} by fixing $Z_k^c=1$. We follow the analysis of the 
ghost-improved model of 
\cite{Ghost_1}, where the motivation stems from a QCD 
analogy, $Z_k^c$ playing an important role in the IR theory \cite{Ghost_2, Ghost_3, Ghost_4, Ghost_5}, 
and the computation revealing 
the interplay between gravitational beta-functions and ghosts.

The ansatz for the scale dependent effective action is
\begin{align}
\label{ghost_ansatz}
&\Gamma_k[g,C,\bar{C};\bar{g},c,\bar{c}] \nonumber  =  \\
&\Gamma_k^{\text{grav}}[g] \, + \, \Gamma_k^{\text{gf}}[g;\bar{g}] \, + \, \Gamma_k^{\text{gh}}[g,C,\bar{C};\bar{g},c,\bar{c}]\,,
\end{align}
where $C, \bar{C}$ are the classical ghost fields, and $c, \bar{c}$ are their associated background 
fields: one has $g_{\mu \nu} = \bar{g}_{\mu \nu} + h_{\mu \nu}$, $\bar{C}_\mu = \bar{c}_\mu + \bar{f}_\mu$
and $C_\mu = c_\mu + f_\mu$, where $h_{\mu \nu}$, $f_\mu$, $\bar{f}_\mu$ mark the expectation 
value of the quantum fluctuations around the background. The ansatz is constructed from the 
$\Gamma_k^{\text{grav}}$ gravitational term, the $\Gamma_k^{\text{gf}}$ gauge-fixing term and the 
$\Gamma_k^{\text{gh}}$ ghost term, with their explicit forms also given in
\cite{Ghost_1}. The harmonic gauge choice is also used in order to compare the result to
the EH truncated gravity without ghost-improvement. Beta-functions in $d$ dimensions are
\begin{align}
    \beta_g =& \, (d-2+ \eta_N) \,g_k \,,\\
    \beta_\lambda =& - (2 - \eta_N) \lambda_k + \frac{1}{2} g_k (4\pi)^{1 - d/2} \nonumber\\
&\Big[
2d(d+1) \Phi^{1,0}_{d/2}(-2\lambda_k) 
- 8d \Phi^{1,0}_{d/2}(0) \nonumber \\
&- d(d+1)\eta_N \tilde{\Phi}^{1,0}_{d/2}(-2\lambda_k) 
+ 4d\eta_c \tilde{\Phi}^{1,0}_{d/2}(0)
\Big], \nonumber
\end{align}
with threshold functions $\Phi_n^{p,q}(w)$, $\tilde{\Phi}_n^{p,q}(w)$, \app{app1}. 

The wave-function renormalization of ghosts 
$Z_k^c$ is completely determined by $g_k$ and $\lambda_k$, since it enters the beta-function 
through the ghost anomalous dimension $\eta_c = -\partial_t \, \text{ln} \, Z_k^c$.

\section{AS gravity at finite temperature}
In order to generalize the $T=0$ gravitational models discussed in the previous sections to their 
finite temperature counterpart, one has to extend the zero temperature RG method to finite temperature. 
The formulation is done in Euclidean spacetime, 
which can be achieved by Wick rotation, i.e., $t \to -it_E$, where $t_E$ is the Euclidean time. This transforms the 
action as $ \int_0^t dt\int d\,^{d-1}x \,\mathcal{L}  \to \int_0^\beta dt_E \int d\,^{d-1} x \, \mathcal{L} (t \to -it_E)$. 
Bosonic fields obey periodic boundary conditions, with $\beta = it = 1/T $ periodicity. Finite temperature QFT 
requires the modification of the momentum integral as
\beq
\label{matsubara}
\int \frac{d\,^d p}{(2 \pi)^d} \hspace{2 mm} \to \hspace{2 mm} T \sum_{m} \int \frac{d\,^{d-1} p}{(2 \pi)^{d-1}} 
\eeq
with the Matsubara summation over $m \in \mathbb{Z}$ exchanging one of the momentum integrals.
This implies that one has to replace the zeroth component of the momentum with $\omega_m$
Matsubara frequencies, 
i.e., $p^2 \to p^2 + \omega_m^2$, where $\omega_m = 2\pi m T$ holds for bosonic frequencies.

Due to the periodic boundary condition, the finite temperature QFT is described on a cylindrical spacetime, 
with radius $R=1/T$. For the reasons discussed in detail in the Introduction, we chose to link $T$ to the scale 
$k$ as $T= \tau k$.

Introducing the Matsubara summation (\ref{matsubara}) and implementing our temperature relation
$T=\tau k$ 
implies the following changes to the the zero-temperature threshold functions (see \app{app1} for details):
\begin{align}
\label{threshold_func_1_T}
\Phi^p_n(w,\tau) = \frac{ \, 2\tau \sqrt{\pi} }{\Gamma\left( n-\frac{1}{2} \right)}\ 
\sum_{m \ = \  -\infty}^{\infty} \int^{\infty}_{0} dy \ y^{n-\frac{3}{2} }  \nn
\frac{R(y) - y \,  R^{\prime}(y)}{\left[ y + (2 m \pi \tau)^2 + R(y) + w \right]^p}, \\
\label{threshold_func_2_T}
\tilde{\Phi}^p_n(w,\tau) = \frac{ \, 2\tau \sqrt{\pi} }{\Gamma\left( n-\frac{1}{2} \right)}\ 
\sum_{m \ = \  -\infty}^{\infty} \int^{\infty}_{0} dy \ y^{n-\frac{3}{2} }  \nn
\frac{R(y)}{\left[ y + (2 m \pi \tau)^2 + R(y) +w \right]^p}.
\end{align}
Subsequently, these expressions are introduced into the beta-functions of the gravitational models,
the only exception being the ghost-improved scenario, in which the generalized
threshold functions are used which 
requires an additional $[y + R(y)]^q \to [y + (2m\pi\tau)^2 + R(y)]^q$ replacement.

\begin{figure}
\begin{center} 
\epsfig{file=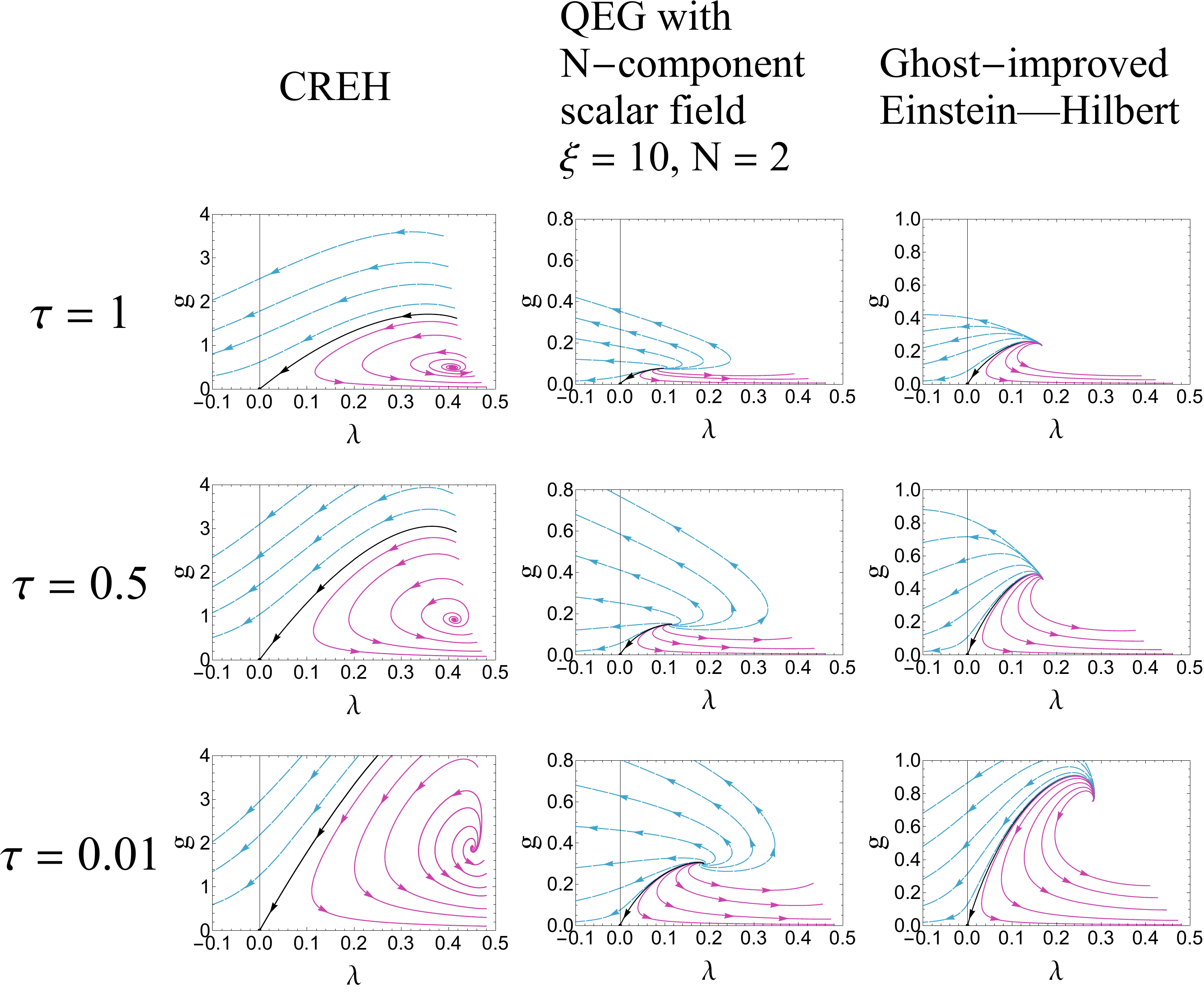,width=8.5cm}
\caption{\label{flowgrid} 
Thermal RG flow diagrams of CREH, QEG with $N$-component scalar field (with $\xi = 10$ and
$N=2$) and Ghost-improved EH gravity for various $\tau$ dimensionless temperature values. With $\tau \to \infty$ 
the $g^* \to 0$ limit is reached in each case.
} 
\end{center}
\end{figure}

\begin{figure}
\begin{center} 
\epsfig{file=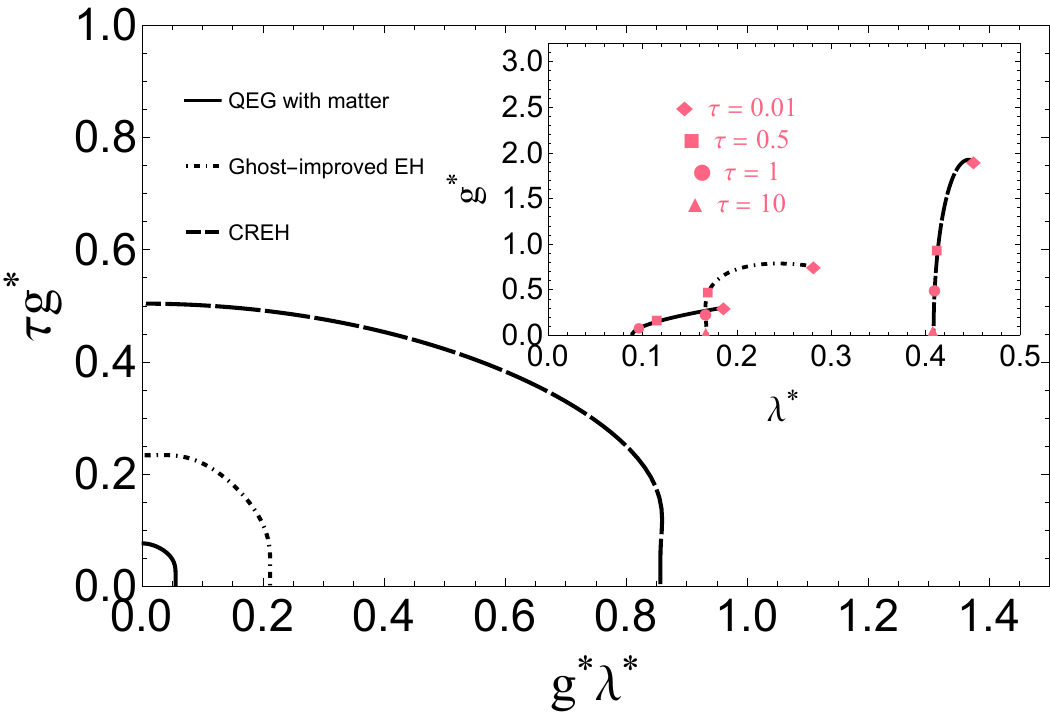,width=8.2cm}
\caption{\label{qptcpt} 
QPT-CPT diagram of various AS gravity models in terms of the dimensionless 
temperature $\tau$ and the $g$-coordinate of the Reuter fixed point. Black lines, i.e., the function 
$g^\star(\tau_c)$, are critical lines which separate the $\lambda<0$ and the $\lambda>0$ phases. 
For a given (but fixed) $g^\star$-value, for $\tau > \tau_c$ or $\tau < \tau_c$ the 
particular model is in the $\lambda<0$ or its $\lambda>0$ phase. The inset shows how the positions of the 
Reuter fixed point ($g^\star$, $\lambda^\star$) changes by $\tau$ in case of the three variants. 
} 
\end{center}
\end{figure}

In all three gravitational models we found that the $g$-component of the Reuter fixed point 
($g^*$) disappears as the temperature increases, i.e., with $\tau \to \infty$, see \fig{flowgrid}.
%
%
Additionally, the slope of the separatrix decreases with increasing $\tau$ in each case and, in the limit
$\tau \to \infty$, only the $\Lambda<0$ phase survives. This result 
shows that AS quantum
gravity naturally predicts a negative cosmological constant for high temperatures. In the early (high-temperature) 
Universe -- whether one considers CREH, QEG with $N$-component scalar field, or ghost-improved 
gravity -- starting from particular initial condition in the $\Lambda <0$ phase \cite{thermal_rg_as_gravity}, 
the decreasing temperature of the Universe drives the system into the $\Lambda>0$ phase, which
is the 
one consistent with current observations, see \fig{qptcpt}.

The system can undergo either a QPT at a fixed finite temperature or a CPT can occur at a fixed quantum parameter. 
There is some freedom in the choice of the parameters, since at $\tau \to \infty$ the $\lambda$-component
of the Reuter fixed point ($\lambda^*$) is essentially constant in all of the discussed models, as seen in the inset 
of \fig{qptcpt}. This implies that the slope of the separatrix $g^*/\lambda^*$, $g^*$ itself, or $g^*\lambda^*$ are all 
adequate candidates for quantum parameters. We chose the latter option, for it is a dimensionless
combination of the couplings 
in $d=4$, i.e., $G_k\Lambda_k = g_k\lambda_k$ applies. Since $g^* \to 0$ is only achieved
in the $\tau \to \infty$ limit, 
it is worth to investigate their combination as the quantum parameter vanishes. As it turns out, $\tau g^*$ 
reaches a constant value in the $g^*\lambda^* \to 0$ limit, which provides the required form for the QPT-CPT diagram 
within all three models, as seen in \fig{qptcpt}.
%
%

In all three models one finds non-computable regions with boundaries signaled by the divergence of 
the anomalous dimension which expands with increasing $\tau$. In CREH gravity, at zero temperature 
this pole appears at $\lambda = 0.5$, however, in the limit $\tau \to \infty$ it is shifted to $\lambda \simeq 0.3$ 
for non-zero $g$ but $\lambda = 0.5$ for vanishing $g$.

\section{Connection to spacetime foliation}
Our modified thermal RG framework shows many formal similarities to RG methods implementing 
foliated spacetimes within the Arnowitt-Deser-Misner (ADM) formalism \cite{ADM_1,ADM_2}, as summarised
in \app{app2}, based on Refs. \cite{ADM_3,Foliation_1_Platania_book, Foliation_2, Foliation_3,Foliation_4}. 

The formulation for the RG equation, where the gravitational degrees of freedom are carried by the ADM fields has been 
constructed in Refs. \cite{Foliation_2,Foliation_3}. In these works the RG flow captured by the ADM-decomposed EH action 
was studied, revealing that the beta-functions parametrically depend on the dimensionless Matsubara (or Kaluza-Klein) 
mass $m$, which is related to the size of the time direction $R$ and it is defined as $m = 2\pi /(Rk)$. If we implement the 
Matsubara formalism $m = 2\pi T /k $ applies, hence the dimensionful temperature takes the form $T = mk/(2 \pi)$. 
Comparing this to our key identification $T=\tau k$, one can see that the dimensionless temperature $\tau$ serves the same 
purpose as the Matsubara mass $m$. Figure 2 of Ref. \cite{Foliation_3} is also
consistent with this observation, as the position 
of the Reuter fixed point exhibits a strikingly similar dependence on $m$ to that shown in the inset of \fig{qptcpt}. The 
beta-function for $m$ was studied in the work mentioned previously, and the $\beta_m(g,\lambda,m) =\partial_tm_k= 0$ 
approximation was taken. Additionally, in \cite{Foliation_5_Horava} both constant and running $m$
were studied in relation to 
Ho\v{r}ava--Lifshitz gravity. This approach might be appropriate regarding foliated spacetimes, but in case of finite temperature 
formalism we advocate for taking $m$, i.e., $\tau$, as constant during the RG transformations. In this way, temperature 
fluctuations are integrated out in a manner similar to quantum fluctuations which makes possible to draw the QPT-CPT 
diagrams of QFT models.

\section{Conclusions}

The main result of this work is a prediction: if one assumes that AS gravity is a viable theoretical framework 
to connect early and late time cosmologies, then based on our thermal RG analysis, the cosmological constant
must be negative in the early and positive in the present Universe. The latter is in agreement with observations 
on the accelerated expansion of the Universe at present while the former needs a verification or falsification.
However, independently of possible future tests on the negative cosmological constant of the early 
Universe, our result receives an important application in string theories which naturally produce the 
cosmological constant with a negative value in agreement with the prediction of this work.

To support the above general statement, in this work we have performed the thermal RG study of three 
variants of the simplest (EH truncated) AS quantum gravity. In all these cases we have found the same 
picture: a negative value for the cosmological constant in the early and positive in the present Universe.
The essence of AS quantum gravity is the presence of the Reuter fixed point whose $g$-coordinate is
vanishing for large temperatures, so thus it results in a negative value for the cosmological constant if 
the the direction of the spiraling RG trajectories around the Reuter fixed point is the same as that of the 
simplest EH truncated model, a requirement fulfilled by the majority of AS quantum gravity models. 

Finally, let us note, our thermal RG scheme predicts a cosmological constant $\Lambda$ with a 
negative value in the early and a positive in the present Universe. This prediction was missing in 
the usual thermal RG literature and it receives important application in the simplest extension of 
the cosmological standard model, i.e., in the so-called $\Lambda_S$CDM model
\cite{lambdaS_cosmo_1, lambdaS_cosmo_2, lambdaS_cosmo_3, lambdaS_cosmo_4, lambdaS_cosmo_5}
which assumes a sign change in $\Lambda$, i.e., switching its sign from a negative to a positive one over 
the time-evolution of the Universe and it has been applied to explain two most relevant problems of the 
$\Lambda$CDM model, i.e., the Hubble-tension \cite{Hubble_1, Hubble_2, Hubble_3} and the 
S$_8$-discrepancy \cite{S8_1, S8_2, S8_3}.

\section*{Acknowledgements} 
The support for the CNR/MTA Italy-Hungary 2023-2025 
Joint Project "Effects of strong correlations in interacting many-body systems and 
quantum circuits" are gratefully acknowledged. 
This work was supported by the University Research Scholarship Programme (EK\"OP-25-2-DE-376) 
of the Ministry of Culture and Innovation and of the National Research, Development and Innovation Fund.

\section*{Note added} 
After the submission several interesting papers appeared  \cite{DDE_5, DDE_6, lambdaS_cosmo_4,lambdaS_cosmo_5,bib0059,bib0060,bib0061}, where the authors discuss observational results in terms of a phenomenological model supporting a scenario compatible with our results. Further investigation on this point is currently on-going.

\appendix

\section{Functional RG study of AS gravity at $T=0$}
\label{app1}
In this Appendix, we provide some details of the $T=0$ temperature functional Renormalization 
Group (RG) flow equations and threshold functions derived for three AS gravity models.

In order to discuss the functional RG study of CREH gravity let us rewrite the Wetterich equation using the 
background field method. In the background field approach the conformal factor of the background metric 
$\bar{g}_{\mu \nu}$ sets the physical scale of $k$. This setting is analogous to the full gravitational RG, with the 
exception that in this case only the conformal factor's quantum fluctuations contribute to the running of the couplings. 
The path integral is taken with respect to the $\chi(x)$ quantum conformal factor field, which can be written as the 
sum of $\chi_B(x)$ fixed background field and $f(x)$ fluctuation, i.e. $\chi = \chi_B + f$. The effective action is 
functionally dependent on $ \bar{f} \equiv \braket{f}$ and $\chi_B$, i.e. $\Ga[\bar{f}; \chi_B]$, or -- using that the 
conformal factor can be given as $\phi \equiv \chi_B + \bar{f}$ -- one can write $\Ga[\phi,\chi_B]$. With the 
background field method the Wetterich RG equation takes the form 
\begin{equation}
k \partial_k \Ga_k[\bar{f}; \chi_B] = \frac{1}{2} \, \text{Tr} \, \frac{k\partial_kR_k[\chi_B]}{\Ga_k^{(2)}[\bar{f}; \chi_B] + R_k[\chi_B]}\,,
\end{equation}
where $R_k[\chi_B]$ is the regulator and $\Ga_k^{(2)}[\bar{f}; \chi_B]$ is the second functional 
derivative with respect to $\bar{f}$ at fixed $\chi_B$. After one inserts an ansatz for the effective 
action $\Ga_k[\bar{f}; \chi_B] =  \int d^4x \, \sqrt{\widehat{g}} {\cal L}^{\rm CREH}_k$ with
\begin{align}
{\cal L}^{\rm CREH}_k = - \frac{3}{4 \pi G_k} \left(-\frac{1}{2} \phi 
\widehat \Box  \phi  + \frac{1}{12} \widehat{\mathcal{R}} \phi^2 - \frac{1}{6} \Lambda_k \phi^4
\right),
\end{align}
into the flow equation and evaluates the trace using the derivative expansion, then compares the terms 
on both sides the beta-functions for the couplings ($G_k$ and $\Lambda_k$) can be derived.

To compute the beta-functions for CREH gravity the following threshold functions are needed
\begin{align}
    \label{threshold}
    \Phi_n^p(w) =& \frac{1}{\Ga(n)} \int_0^\infty dy\, y^{n-1}\, \frac{R(y)-yR'(y)}{[y+R(y)+w]^p}\,, \\
    \tilde{\Phi}_n^p(w) =& \frac{1}{\Ga(n)} \int_0^\infty dy\, y^{n-1}\, \frac{R(y)}{[y+R(y)+w]^p}\,. \nonumber
\end{align}
The anomalous dimension is
\begin{equation}
\label{an_dim}
    \eta_N(k) = \frac{g_k B_1(\lambda_k)}{1-g_kB_2(\lambda_k)} \,,
\end{equation}
where $B_1(\lambda_k)$ and $B_2(\lambda_k)$ functions take the form
\begin{align}
    B_1(\lambda_k) =& \, \frac{1}{6\pi} \, [\, \Phi_1^1(-2\lambda_k) - \Phi_2^2(-2\lambda_k) \,]\,, \\
    B_2(\lambda_k) =& - \frac{1}{12 \pi} \, [\, \tilde{\Phi}_1^1(-2\lambda_k) - \tilde{\Phi}_2^2(-2\lambda_k) \,]\,. \nonumber
\end{align}

In order to discuss the functional RG study of QEG with matter let us compute the anomalous dimension 
for Einstein gravity coupled to $N$-component scalar field. To do this, the functions listed below are required
\begin{align}
    B_1(\lambda_k) =& \, \frac{1}{6 \pi} \, [\,10\,\Phi_1^1(-2\lambda_k) + (N-8) \, \Phi_1^1(0) \, - \\
    &-\, 36 \,\Phi_2^2(-2\lambda_k) - (12+6\,\xi \, N) \,\Phi_2^2(0) \,]\,, \nonumber\\
    B_2(\lambda_k) =& \, \frac{1}{6 \pi}\,[\, 18\,\tilde{\Phi}_2^2(-2\lambda_k) -5 \tilde{\Phi}_1^1(-2\lambda_k) \,]\,. \nonumber
\end{align}

Finally, the generalized threshold functions needed for Ghost-improved EH gravity are
\begin{align}
\Phi_n^{p,q}(w) &= \frac{1}{\Gamma(n)} \int_0^\infty dy\, y^{n-1} \frac{R(y) - y R'(y)}{(y + R(y) + w)^p (y + R(y))^q}, \nonumber\\
\tilde{\Phi}_n^{p,q}(w) &= \frac{1}{\Gamma(n)} \int_0^\infty dy\, y^{n-1} \frac{R(y)}{(y + R(y) + w)^p (y + R(y))^q}, \nonumber\\
\check{\Phi}_n^{p,q}(w) &= \frac{1}{\Gamma(n)} \int_0^\infty dy\, y^{n-1} \frac{R'(y) (R(y) - y R'(y))}{(y + R(y) + w)^p (y + R(y))^q}, \nonumber\\
\hat{\Phi}_n^{p,q}(w) &= \frac{1}{\Gamma(n)} \int_0^\infty dy\, y^{n-1} \frac{R(y) R'(y)}{(y + R(y) + w)^p (y + R(y))^q}.
\end{align}
One can notice that in the case of $\Phi_n^{p,q}(w)$ and $\tilde{\Phi}_n^{p,q}(w)$ the choice $q = 0$ recovers the functions seen in Eqs. (\ref{threshold}).

The anomalous dimensions of the Newton constant and the ghost wave-function renormalization are given as
\begin{align}
\eta_N &= \frac{g B_1 + g^2 \left( C_3 C_4 - B_1 C_2 \right)}%
{1 - g \left( B_2 + C_2 \right) + g^2 \left( B_2 C_2 - C_1 C_3 \right)}, \\[10pt]
\eta_c &= \frac{g C_4 + g^2 \left( B_1 C_1 - B_2 C_4 \right)}%
{1 - g \left( B_2 + C_2 \right) + g^2 \left( B_2 C_2 - C_1 C_3 \right)}, 
\end{align}
with the functions 
\begin{align}
B_1(\lambda) &= \frac{1}{3} (4\pi)^{1 - \frac{d}{2}} \Big[ 
d(d + 1)\, \Phi^{1,0}_{\frac{d}{2} - 1} 
- 6d(d - 1)\, \Phi^{2,0}_{\frac{d}{2}} 
\nonumber\\ 
&- 4d\, \Phi^{0,1}_{\frac{d}{2} - 1} 
- 24\, \Phi^{0,2}_{\frac{d}{2}} 
\Big], \nonumber\\
B_2(\lambda) &= -\frac{1}{6} (4\pi)^{1 - \frac{d}{2}} \left[
d(d + 1)\, \tilde{\Phi}^{1,0}_{\frac{d}{2} - 1} 
+ 6d(d - 1)\, \tilde{\Phi}^{2,0}_{\frac{d}{2}} 
\right], \nonumber\\
C_1(\lambda) &= (4\pi)^{1 - \frac{d}{2}} \left[ 2 C_{\text{gr}}\, \tilde{\Phi}^{2,1}_{\frac{d}{2} + 1} - 4d \left( \tilde{\Phi}^{2,2}_{\frac{d}{2} + 2} + \hat{\Phi}^{2,2}_{\frac{d}{2} + 2} \right) \right], \nonumber\\
C_2(\lambda) &= (4\pi)^{1 - \frac{d}{2}} \left[ 2 C_{\text{gh}}\, \tilde{\Phi}^{1,2}_{\frac{d}{2} + 1} + 4d \left( \tilde{\Phi}^{2,2}_{\frac{d}{2} + 2} + \hat{\Phi}^{2,2}_{\frac{d}{2} + 2} \right) \right], \nonumber\\
C_3(\lambda) &= \frac{1}{3} (4\pi)^{1 - \frac{d}{2}} \left[ 2d\, \tilde{\Phi}^{0,1}_{\frac{d}{2} - 1} + 12\, \tilde{\Phi}^{0,2}_{\frac{d}{2}} \right], \nonumber\\
C_4(\lambda) &= - (4\pi)^{1 - \frac{d}{2}} \left[ 4 C_{\text{gr}}\, \Phi^{2,1}_{\frac{d}{2} + 1} + 4 C_{\text{gh}}\, \Phi^{1,2}_{\frac{d}{2} + 1} \right].
\end{align}
The coefficients $C_{\text{gr}}$ and $C_{\text{gh}}$ take the form
\begin{align}
C_{\text{gr}} = \frac{4d^2 - 9d - 2 }{d - 2}, \hspace{5 mm}
C_{\text{gh}} = \frac{2d^2 - 5d - 2}{d - 2}\,.
\end{align}
One can recover the Einstein-Hilbert truncated anomalous dimension without ghost-improvements by setting $C_i = 0$.

\section{Arnowitt-Deser-Misner formalism}
\label{app2}
The Arnowitt-Deser-Misner (ADM) formalism is used to connect spacetime foliation and thermal RG.
In general relativity the metric is equipped with Lorentzian signature, however $\Gamma_k$ is given by an Euclidean 
path integral thus the Lorentzian spacetime has to be recovered by Wick-rotation. QFT calculations are usually 
done on a fixed Minkowski background which provides a clear notion of time. However, when discussing 
dynamical spacetimes, the role of time can be questioned. To address this the ADM-formalism is used, in which the 
spacetime metric $g_{\mu\nu}$ is decomposed into a lapse function $N_l$, a shift-vector $N_i$ and a metric on spatial 
slices $\sigma_{ij}$. These are needed, because the $4D$ spacetime is sliced into $3D$ spatial hypersurfaces with metric 
$\sigma_{ij}$, each hypersurface labeled with time parameter $t$. $N_l$ -- being related to the separation between 
hypersurfaces -- and $N_i$ -- which is a displacement related to a point passing to the next surface -- describe how 
to weld the hypersurfaces together to form the foliation, thus imprinting the Euclidean spacetime with a distinguished 
direction. The resulting preferred time direction enables the computation of transition amplitudes from an initial to a final slice.


\end{document}